\documentclass[12pt,preprint]{aastex}

\shorttitle{Virtual Reality: A Definition History}
\shortauthors{S. Bryson}

\begin{document}

\title{Virtual Reality: A Definition History - A Personal Essay}

\author{
Steve~Bryson\altaffilmark{1},
}

\altaffiltext{1}{NASA Ames Research Center, Moffett Field, CA 94035; steve.bryson@nasa.gov}
%

\begin{abstract}

This essay, written in 1998 by an active participant in both virtual reality development and the virtual reality definition debate,
discusses the definition of the phrase ``Virtual Reality'' (VR).  I start with history from a personal perspective,
concentrating on the debate between the ``Virtual Reality'' and ``Virtual Environment'' labels in the late 1980's and early 1990's.  
Definitions of VR based on specific technologies are shown to be unsatisfactory.
I propose the following definition of VR, based on the
striking effects of a good VR system: ``Virtual Reality is the use of computer technology to create the effect of
an interactive three-dimensional world in which the objects have a sense of
spatial presence.''  The justification for this definition is discussed in detail, and is favorably compared with the
dictionary definitions of ``virtual'' and ``reality''.  The implications of this definition for virtual reality technology 
are briefly examined.

\end{abstract}


\section{Preface} \label{section:preface}
I wrote the essay below in 1998 in response to a discussion of the definition of virtual reality in a virtual reality interest group
run by Linda Jacobsen.  
This essay found its way into an online ``Omnibus Lexicon Definition Supplement'' on a now 
defunct website by the Fourth Wave Group, under the title ``Virtual Reality: A Definition History''.  
Because this web site is no longer available, and there are at least 13 citations to this essay in the literature
according to Google Scholar,
I am providing this essay here.  This version is lightly edited from the original for clarity and to remove discussions specific to the
virtual reality interest group.  Major new comments are inserted in square brackets: [].

It is my opinion that the definition of virtual reality presented in this essay has stood the test of time.


\section{Virtual Reality: A Definition History (written in 1998)} \label{section:essay}
I would like to add to the discussion on definitions in virtual reality (VR), as my position
on this has evolved through more than a decade of watching, and participating in, the ``VR
definition wars''.

\subsection{History}

First I'd like to review some of the history.  If I can be allowed a moment
of nostalgia, I fondly recall sometime in 1985 or '86 talking to Jaron
Lanier on the phone (I worked for him at the time at VPL Research), and him
telling me about the lunch he'd had that day with Scott Fisher then of the
NASA Ames Virtual Interactive Environment Workstation (VIEW) lab.  
[The VIEW lab was one of the first multi-modal virtual reality implementations,
incorporating wide-field, head-tracked stereoscopic display, hand and gesture tracking
and three-dimensional sound rendering supported by 30 Hz three-dimensional rendering
and workstation-class computation.]
They were talking about what to call what Scott was
doing in the VIEW lab when Jaron coined the term ``virtual reality''.  Jaron
told me that he thought it was an accurate description of what Scott was
doing.  Scott and others at, {\it e.g.} the University of North Carolina, had already been calling it virtual
environments (VE): VIEW stands for Virtual Interactive Environment Workstation,
and was the follow-on to Michael McGreevey's Virtual Visual
Environment Display (VIVED) project at NASA Ames \citep{VIVED}.  

I'm happy to say that I joined Scott's VIEW lab
in 1988.

For the next several years there was a battle between the ``virtual reality'' 
and the ``virtual environment'' camps over appropriate
nomenclature, with the academics tending towards VE and the
hacker/commercial/press sectors tending towards VR.  
[The VE side generally expressed that the phrase ``virtual reality'' was too vague
and somewhat oxymoronic.  Though I tended to agree with this judgement at the 
time, as described in \ref{section:def} I have since decided that ``virtual reality'' is actually precise and not at all 
oxymoronic.] At about this time the
phrase ``artificial reality'' was popular in Japan.  VR became generally
accepted in the 92-93 time frame, as signified by: 
\begin{itemize}
\item the creation in 1992 of
the National Academy of Sciences/National Research Council Committee on
Virtual Reality Research and Development (though the resulting 1995 NAS/NRC
report \citep{NRC} was ambivalent on the VR/VE issue),
\item the creation of the 
Virtual Reality Annual International Symposium (VRAIS)
conference and the IEEE Symposium on Virtual Reality Research and
Development in 1993,
\item perhaps most importantly, Fred Brooks finally
agreeing to the VR label.  
\end{itemize}
\noindent I feel that VR won by sheer volume: every time
we said ``VE'' we had to explain that we meant ``VR'' as everyone had heard of
VR even if they didn't know what it meant.  As I'll explain below I've
completely come to peace with the VR label through the analysis of definitions.

In the meantime there were very many attempts to define VR/VE.  
Agreement on definition was
elusive.  We all knew that what we were doing was special, but expressing
just what was special proved surprisingly difficult.  Most of these
attempts tried to define VR through the technology used to achieve this
unique effect, but this approach tended to lead to debates over whether
this or that technology was ``required'' for VR.  In the meantime it seemed
that the point was lost, and in any case such a definition is uninformative
to those who have not themselves experienced the technology in action.
Simply saying ``head tracking'' does not convey the power of a head-tracked
display in action.

In the meantime VR was being used to describe everything from
three-dimensional photo-realistic rendering through non-graphical
simulation to artificial intelligence.  And this does not even count what
was happening in literary and artistic circles where ``virtual reality'' was
being used in very creative ways.  It was clear that Jaron had picked a
phrase which ignited something in people, though it is not clear that Jaron
intended the effect to be quite this broad.  It became apparent that VR was
in danger of being so broadly used as to lose meaning entirely.  This was
one of the primary arguments against VR used by the VE camp.  So it became
critical to many of us to come up with a good definition of VR.  By this
time defining it in terms of technology was completely untenable, because those
who were using the term more broadly could simply disagree.

\subsection{A Definition of Virtual Reality}\label{section:def}

It was clear that we had to base our definition on the effects of VR: after
all it was primarily the (promised) effects, and only secondarily the
technology which seemed to make people most excited.  Where I ended up was
defining VR in terms of its cognitive effects: creating a sense of spatial
presence, possibly a sense of immersion, a sense of interaction with
objects.  We then had to clarify the difference between VR and
telepresence, where remotely sensed objects are presented to our senses.
Combining these observations I have settled on the following definition:

\noindent {\bf Virtual Reality is the use of computer technology to create the effect of
an interactive three-dimensional world in which the objects have a sense of
spatial presence.}

\noindent I want to examine this definition piece by piece:

\begin{itemize}
\item ``Computer technology'' is required to distinguish VR from telepresence and
other remote sensing approaches.  This requirement is driven by the ability
to use computer programs to create interesting and novel tailor-made
environments, which is where I think a lot of the popular interest in VR
comes from.

\item ``Effect'' rather than ``illusion'' because I feel that it is a cognitive
effect that is achieved (more on this later), rather than an illusion.
Saying ``effect'' rather than ``illusion'' also undercuts the presumption of
``fooling'' the user.

\item ``Interactive'' to distinguish VR from conventional animation.  I also
think that it is the ability to interact with the virtual world which is
behind much of the popular excitement.

\item ``Three-dimensional world'' to exclude text-based environments and to limit
discussion away from 1D and 2D programs (which would encompass essentially
all of interactive computer graphics).

\item ``Objects have a sense of spatial presence'' means that the objects seem to
have a spatial location independent of both the user and the display
technology.  This is, I feel, at the heart of what is special about VR.
\end{itemize}

\noindent I shorten this definition to the buzz phrase ``interaction with things, not
(possibly animated) pictures of things.''

By now I'm sure you've noticed the absence of ``immersion'' in the
definition.  I take ``immersion'' to mean ``being surrounded'', and so a sense
of immersion would require a sense that I am surrounded by the environment:
things can be behind me and if so I'll see them when I turn around.
Initially my definition included immersion, but having worked with such
non-immersive displays such as the responsive workbench \citep{immersiveWorkbench} 
or the immersadesk \citep{immersadesk},
it is clear to me that all the power of VR can happen without any sense of
immersion whatsoever.  So I dropped ``immersion'' as a requirement.

Also absent from this definition is a requirement that the virtual
environment mimic the real world either in terms of content or in terms of
interface.  I feel that such requirements miss the point and tend to limit
the creative development of effective virtual environments.

This definition is tentative, but is the best I've been able to do to date.

In the meantime I was meditating on the phrase ``virtual reality'' itself.
Most people react to it as an oxymoron but I always had a sense that Jaron
was exactly right in choosing this phrase.  So one day I went to several
dictionaries and looked up ``virtual'' and reality''  There were several
definitions of each, but the ones that stood out to me as appropriate to
our purposes were the following:
\begin{itemize}
\item Virtual: to have the effect of being such without actually being such
\end{itemize}
\noindent and
\begin{itemize}
\item Reality: the property of being real
\item Real: the property of having concrete existence.
\end{itemize}
\noindent Putting these together, ``Virtual Reality'' means ``to have the effect of
having concrete existence without actually having concrete existence''.  I
think this is  an impressively accurate description of what is special
about what we are doing in VR.

Now of course once you accept the above definition of virtual reality, it
has real implications about the technology required.  When your display is
visual or auditory, head tracking is clearly required to maintain a sense
that objects have an independent spatial presence as you move your head.  
The desired effect
is that the virtual object stays put (assuming it's not moving) as you move
around.  My catchphrase for this when using a visual or auditory display
is: ``if you move your head and nothing happens it ain't VR''.  Whatever your
modality, ``near-real-time'' (or whatever you want to call it) performance is
again required for spatial presence.  Also the relationship between the
user and virtual objects is critical to providing the effect of spatial
presence, so something about the user must be tracked in all cases.  Note
that visual fidelity is not required (for a visual display) in this
definition: the 100x100 monochrome display in the 1988 VIEW lab had
terrible fidelity (not to mention Ivan Sutherland's original display from
1967), but it provided a strong sense of spatial presence which deeply
impressed those who tried it.

The above helps clarify the problem with definitions based on technology: I
have yet to see a technology requirement beyond ``near-real-time''
performance and user tracking which must be in every system anyone can
mention which clearly (intuitively) is a VR system.  But clearly not all
fast systems which track the user are VR systems: the results of the
tracking can be used in a way that have nothing to do with spatial presence
or immersion.

\clearpage

\end{document}